\documentclass[12pt]{article}

\begin{document}
\pagenumbering{arabic}
\begin{titlepage}
\title{Teleparallel gauge theory of gravity}

\author{J. W. Maluf$\,^{(1,a)}$ and F. F. Faria$\,^{(2,b)}$}
\date{}
\maketitle

{\footnotesize
\noindent{  (1) Instituto de F\'{\i}sica, Universidade de 
Bras\'{\i}lia, C.P. 04385, 70.919-970 Bras\'{\i}lia DF, Brazil}\par
\bigskip
\noindent{ (2) Instituto de F\'{i}sica, Universidade Estadual do Piau\'{i},
Campus Poeta Torquato Neto, 64.002-150 Teresina PI, Brazil}}

\begin{abstract}
In this work a tetrad theory of gravity, invariant under conformal 
transformations, is investigated. The action of the theory is similar to the 
action of Maxwell's electromagnetism. The role of the electromagnetic
gauge potential is played by the trace of the torsion tensor of the 
Weitzenb\"ock spacetime. It is shown that all static, spherically symmetric 
space-times, are solutions of the vacuum field equations. However, by fixing 
the gauge in the linearized form of the vacuum field equations, the usual 
Newtonian limit for the gravitational field is obtained.
\end{abstract}
\vfill
\noindent PACS numbers: 04.20.-q, 04.20.Cv, 04.50.Kd\par
\bigskip
\noindent (a) wadih@unb.br, jwmaluf@gmail.com\par
\noindent (b) felfrafar@hotmail.com\par
\end{titlepage}
\bigskip
\section{Introduction}
Conformal invariance is an important symmetry in some attempts to quantize 
gravity \cite{tHooft1,tHooft2}. The symmetry is normally realized in 
alternative formulations of general relativity, which modify both the small
and large scale descriptions of the space-time geometry. The behaviour of the
gravitational field at very small distances has to be modified in order to
construct a renormalizable and unitary theory of quantum gravity. At large
scales, the modified theory has to yield an acceptable solution to the 
dark matter and dark energy problems. Attempts based on the conformally 
invariant Weyl theory, which is quadratic in the curvature tensor, have been
recently carried out in the literature \cite{Mannheim1,Mannheim2,Moon2}.

The usual procedures for modifying the standard formulation of general 
relativity and obtaining a theory endowed with conformal invariance 
are the following. Either one adds a scalar field to the 
Hilbert-Einstein action integral, together with a suitable kinetic term for 
the scalar field, or one considers alternative theories like the quadratic 
Weyl theory. The latter is known to be the only {\it metrical} theory
that exhibits conformal invariance. However, it is not the only 
{\it geometrical} theory. It is possible to construct an infinity of 
geometrical theories of gravity out of the tetrad field, which are invariant
under conformal transformations, and which may play a role in the formulation
of the quantum theory of gravity. Teleparallel theories of gravity, endowed 
with conformal invariance, have been recently investigated \cite{Maluf1}.
It was found that there is an infinity of theories, beyond the 
quadratic Weyl theory, that display conformal invariance.

In this article, we address a particular theory of gravity that, besides 
conformal invariance, displays three interesting features: (i) the functional
structure of the theory is similar to Maxwell's theory of electromagnetism;
(ii) all static, spherically symmetric geometries, including the Schwarzschild
geometry, are solutions of the vacuum field equations; and (iii) by fixing 
the gauge in the linearized form of the field equations, we obtain the usual
Newtonian limit. The emergence of the Schwarzschild solution in the context of
a Maxwell type theory of gravity is an intriguing result. This theory can be
understood as a teleparallel theory for the gravitational field. 

In Sect. 2 we review the construction of conformally invariant 
teleparallel theories of gravity, and in Sect. 3 we present the Maxwell type
theory, the spherically symmetric solutions and the linearized solution of the
vacuum field equations. Finally, in Sect. 4 we present our conclusions.

\bigskip
Notation: space-time indices $\mu, \nu, ...$ and SO(3,1) indices $a, b, ...$
run from 0 to 3. Time and space indices are indicated according to
$\mu=0,i,\;\;a=(0),(i)$. The tetrad field is denoted $e^a\,_\mu$, and the 
torsion tensor reads $T_{a\mu\nu}=\partial_\mu e_{a\nu}-\partial_\nu e_{a\mu}$.
The flat, Minkowski spacetime metric tensor raises and lowers tetrad indices 
and is fixed by $\eta_{ab}=e_{a\mu} e_{b\nu}g^{\mu\nu}= (-1,+1,+1,+1)$. The 
determinant of the tetrad field is represented by $e=\det(e^a\,_\mu)$.

The space-time geometry is defined here by the tetrad field only, and the only
possible non-trivial definition for the torsion tensor is given by 
$T^a\,_{\mu\nu}$. This torsion tensor is related to the antisymmetric part of 
the Weitzenb\"ock  connection 
$\Gamma^\lambda_{\mu\nu}=e^{a\lambda}\partial_\mu e_{a\nu}$, which establishes
the Weitzenb\"ock spacetime. The curvature of the Weitzenb\"ock connection 
vanishes. However, the tetrad field also yields the metric tensor, which 
determines the Riemannian geometry. Thus, in the framework of a 
geometrical theory based on the tetrad field, one may use the concepts of 
both Riemannian and Weitzenb\"ock geometries.

\section{Teleparallel theories with conformal ivariance}

In this section we review the results recently obtained in \cite{Maluf1}.
A conformal transformation on the space-time metric tensor  
transforms $g_{\mu\nu}$ into $\tilde{g}_{\mu\nu}=e^{2\theta(x)}\,g_{\mu\nu}$, 
where $\theta(x)$ is an arbitrary function of the space-time coordinates. The 
conformal transformations on the tetrad field and on its inverse are defined 
by

\begin{equation}
\tilde{e}_{a\mu}=e^{\theta(x)}\,e_{a\mu}\,, \ \ \ \ \ \ \ \ \ \ \
\tilde{e}^{a\mu}=e^{-\theta(x)}\,e^{a\mu}\,.
\label{1}
\end{equation}
The transformation of the projected components of the torsion tensor 
$T_{abc}=e_b\,^\mu e_c\,^\nu (\partial_\mu e_{a\nu}-\partial_\nu e_{a\mu})$ is
straightforward. It is given by

\begin{eqnarray}
\tilde{T}_{abc}&=&
e^{-\theta}(T_{abc} + \eta_{ac}\,e_b\,^\mu \partial_\mu \theta
-\eta_{ab}\,e_c\,^\mu \partial_\mu \theta)\,, \nonumber \\
\tilde{T}^{abc}&=&
e^{-\theta}(T^{abc} + \eta^{ac}\,e^{b\mu} \partial_\mu \theta
-\eta^{ab}\,e^{c\mu} \partial_\mu \theta)\,. 
\label{2}
\end{eqnarray}
As a consequence of the equation above, the trace of the torsion tensor 
$T_a=T^b\,_{ba}$ transforms as 

\begin{eqnarray}
\tilde{T}_a&=&e^{-\theta}(T_a-3\,e_a\,^\mu \partial_\mu \theta)\,, 
\nonumber \\
\tilde{T}^a&=&e^{-\theta}(T^a-3\,e^{a\mu} \partial_\mu \theta)\,.
\label{3}
\end{eqnarray}
We also have 

\begin{equation}
\tilde{T}_\mu= T_\mu -3\partial_\mu \theta\,,
\label{4}
\end{equation}
where $T_\mu=T^\lambda\,_{\lambda\mu}$.

With the help of equations (2) and (3), it is possible to verify that the 
behaviour of the three quadratic terms that determine the Lagrangian density 
of the standard teleparallel theories of gravity is given by

\begin{eqnarray}
\tilde{T}^{abc}\tilde{T}_{abc}&=& e^{-2\theta}(T^{abc}T_{abc}-
4T^\mu \partial_\mu\theta+6g^{\mu\nu}\partial_\mu\theta\partial_\nu \theta)\,,
\nonumber \\
\tilde{T}^{abc}\tilde{T}_{bac}&=& e^{-2\theta}(T^{abc}T_{bac}-
2T^\mu \partial_\mu\theta+3g^{\mu\nu}\partial_\mu\theta\partial_\nu \theta)\,,
\nonumber \\
\tilde{T}^a\tilde{T}_a&=&e^{-2\theta}(T^a\,T_a -6T^\mu \partial_\mu \theta
+9g^{\mu\nu}\partial_\mu\theta\partial_\nu \theta)\,.
\label{5}
\end{eqnarray}
In view of the equations above, it is straightforward to check that the quantity

\begin{equation}
L={1\over 4}T^{abc}T_{abc}+{1\over 2}T^{abc}T_{bac}-{1\over 3}T^aT_a\,
\label{6}
\end{equation}
transforms under a conformal transformation according to 
$\tilde{L}=e^{-2\theta} L$. As a consequence of Eq. (1), we find that for the 
determinant $e$ of the tetrad field we have $\tilde{e} = e^{4\theta}\,e$. 
Therefore, by introducing a scalar field $\phi$ that is assumed to transform as

\begin{equation}
\tilde{\phi}=e^{-\theta} \phi\,,
\label{7}
\end{equation}
it is easy to verify that 

\begin{equation}
e\phi^2\biggl( 
{1\over 4}T^{abc}T_{abc}+{1\over 2}T^{abc}T_{bac}-{1\over 3}T^aT_a \biggr)
\label{8}
\end{equation}
is invariant under coordinate transformations and conformal transformations
\cite{Maluf1}. It follows from Eq. (4) that a covariant derivative for the 
scalar field may be defined, 

\begin{equation}
D_\mu \phi=\biggl(\partial_\mu -{1\over 3} T_\mu \biggr)\phi\,.
\label{9}
\end{equation}
It is easy to verify that $\tilde{D}_\mu \tilde{\phi}=e^{-\theta}\,D_\mu \phi$.

In view of equations (5-9), it is possible to construct two sets of 
Lagrangian densities that are invariant under conformal transformations, as
described in \cite{Maluf1}. The first set is a one-parameter family of 
theories constructed out of the tetrad field and of the scalar field, and is 
given by

\begin{equation}
{\cal L} = k e\biggl[
-\phi^2\biggl( {1\over 4}T^{abc}T_{abc}+
{1\over 2}T^{abc}T_{bac}-{1\over 3}T^aT_a \biggr) 
+k'g^{\mu\nu}D_\mu \phi D_\nu \phi\biggr]\,,
\label{10}
\end{equation}
where $k=1/(16\pi G)$, and $k'$ is an arbitrary constant parameter. By fixing
$k'=6$, we arrive at the teleparallel equivalent of general relativity
\cite{Maluf1}. The second set is a four-parameter family of theories, and reads

\begin{equation}
{\cal L}(e_{a\mu})= e  L_1 L_2 \,,
\label{11}
\end{equation}
where

\begin{equation}
L_1=A\, T^{abc}T_{abc}+ B\, T^{abc}T_{bac}+ C\, T^aT_a\,,
\label{12}
\end{equation}

\begin{equation}
L_2=A'\, T^{abc}T_{abc}+ B'\,T^{abc}T_{bac}+ C'\, T^aT_a\,.
\label{13}
\end{equation}
The constant coefficients in the expressions above are required to satisfy

\begin{equation}
2A+B+3C=0\,, \ \ \ \ \ \ \ \ \ \ 2A' +B' +3C'=0\,.
\label{14}
\end{equation}
The field equations derived from the Lagrangian density (11) are rather
intricate, and it is not possible to envisage any simple solution of this
four-parameter theory. 

\section{A Maxwell-type theory of gravity}

A conformally invariant teleparallel theory of gravity, that was not noticed 
in the analysis of \cite{Maluf1}, is a theory whose Lagrangian density is
exactly similar to Maxwell's theory of electromagnetism. It is constructed out
of the tetrad field only, and is given by

\begin{equation}
{\cal L} = -{1\over 4}e\,
g^{\mu\alpha} g^{\nu\beta}F_{\mu\nu}F_{\alpha\beta}\,,
\label{15}
\end{equation}
where

\begin{equation}
F_{\mu\nu}=\partial_\mu T_\nu -\partial_\nu T_\mu\,,
\label{16}
\end{equation}
and $T_\mu=T^\lambda\,_{\lambda\mu}$. Under transformation  
(4), we have $\tilde{F}_{\mu\nu}= F_{\mu\nu}$, and therefore ${\cal L}$ given
by Eq. (15) is invariant under conformal gauge transformations. The constant 
factor $-1/4$ is introduced just to emphasize the similarity with Maxwell's 
theory.

The identification of a gauge field with the trace of the torsion tensor is not
a novelty. Dirac \cite{Dirac} and Utyiama \cite{Utyiama} have already
considered a general affine connection such that this feature takes place. 
In these attempts, the theories are formulated with metric and torsion as 
independent field quantities, and with the addition of a scalar field
(see also refs. \cite{Poplawski,Maluf2,Obukhov1,Obukhov2,Nieh,Dereli} and 
references therein; these approaches are understood as metric affine
theories of gravity, which have been thoroughly investigated by Hehl 
{\it et. al.} \cite{Hehl}). In contrast, in the present 
analysis the only field variable is the tetrad field.

The vacuum field equations obtained by varying the action integral constructed
out of (16) read

\begin{eqnarray}
&{}&e_{a\sigma}e_{b\mu}\partial_\lambda 
[\partial_\nu (eF^{\sigma\nu})e^{b\lambda}]
-e_{a\lambda}e_{b\mu}\partial_\sigma 
[\partial_\nu (eF^{\sigma\nu})e^{b\lambda}] \nonumber \\
&+&\partial_\nu (e F^{\lambda\nu})T_{a\mu\lambda}-e{\cal T}_{a\mu}=0\,,
\label{17}
\end{eqnarray}
where

\begin{equation}
{\cal T}_{a\mu}=F^\lambda\,_aF_{\lambda\mu}-
{1\over 4} e_{a\mu}F_{\alpha\beta}F^{\alpha\beta}\,.
\label{18}
\end{equation}
Expression (18) is similar to the standard energy-momentum tensor for the 
electromagnetic field. Although the theory investigated here is considered a
theory for the gravitational field only, we may of course bring to the present
context insights from the standard electromagnetic theory.
We mention that a theory constructed out of a term similar to $F_{\mu\nu}$
has been investigated in \cite{Hammond1}. The theory in the later
reference is constructed out of the torsion and Riemann tensors, and yield
second order field equations, in contrast to the present approach. We also
remark that the gauge transformation given by Eq. (9) is equivalent to the 
transformation of the torsion tensor (Eq. (14a)) of \cite{Hammond2},
where quadratic theories of gravity with second order field equations were
addressed. 

We will show that all static, spherically symmetric space-time geometries are 
solutions of the vacuum field equations. Let us consider the line element 

\begin{equation}
ds^2=-A^2dt^2+B^2dr^2 +r^2d\theta^2 +r^2 \sin^2\theta d\phi^2\,,
\label{19}
\end{equation}
where $A(r)$ and $B(r)$ are arbitrary functions of the radial coordinate $r$.
The set of tetrad fields adapted to stationary observers in space-time, and
that yields the line element (19), is given by

\begin{equation}
e_{a\mu}=\pmatrix{-A&0&0&0\cr
0&B\sin\theta\cos\phi& r\cos\theta\cos\phi&-r\sin\theta\sin\phi\cr
0&B\sin\theta\sin\phi& r\cos\theta\sin\phi& r\sin\theta\cos\phi\cr
0&B\cos\theta&-r\sin\theta&0}\,.
\label{20}
\end{equation}
Out of Eq. (20) we obtain the following non-vanishing components of 
$T_{a\mu\nu}$,

\begin{eqnarray}
T_{(0)01}&=& \partial_1 A\,, \nonumber \\
T_{(1)12}&=&(1-B)\cos\theta\cos\phi\,, \nonumber \\
T_{(2)12}&=&(1-B)\cos\theta\sin\phi\,, \nonumber \\
T_{(3)12}&=&-(1-B)\sin\theta\,, \nonumber \\
T_{(1)13}&=&-(1-B)\sin\theta\sin\phi\,, \nonumber \\
T_{(2)13}&=&(1-B)\sin\theta\cos\phi\,.
\label{21}
\end{eqnarray}
The expressions above yield only three non-vanishing components of 
$T_{\lambda\mu\nu}$,

\begin{eqnarray}
T_{001}&=& A\partial_1 A\,, \nonumber \\
T_{212}&=&r(1-B) \,, \nonumber \\
T_{313}&=&r(1-B)\sin^2\theta \,.
\label{22}
\end{eqnarray}
We may then calculate the traces 
$T_\mu=T^\lambda\,_{\lambda\mu}(t,r,\theta,\phi)$. We obtain

\begin{eqnarray}
T_0&=&0 \,, \nonumber \\
T_1&=& -{1\over A} \partial_1 A -{2\over r}(1-B) \,, \nonumber \\
T_2&=&0\,, \nonumber \\
T_3&=&0\,.
\label{23}
\end{eqnarray}
From Eq. (23) we easily obtain

\begin{equation}
F_{\mu\nu}(t,r,\theta,\phi)=0\,,
\label{24}
\end{equation}
and therefore we conclude that Eq. (20) is a solution of Eq. (17), for
arbitrary functions $A(r)$ and $B(r)$. The Schwarzschild metric tensor is 
obtained by identifying $A^2=(1-2m/r)$ and $B^2=(1-2m/r)^{-1}$. 

By establishing a weak field approximation for the tetrad field according to

\begin{equation}
e^a\,_\mu  \simeq \delta^a_\mu + {1\over 2} h^a\,_\mu\,,
\label{25}
\end{equation}
the field equations (17) may be reduced to a linearized form, to first order 
in $h^a\,_\mu$ (in contrast to the field equations derived from the Lagrangian
density (11); the lowest order field equation obtained from the latter is of 
the order $(h_{a\mu})^2\;$). By imposing Eq. (25) to (17), and transforming 
all indices into space-time indices, we obtain

\begin{equation}
\partial_\mu\partial_\nu\partial^\sigma\partial_\rho h^{\rho\nu}-
\partial_\mu\partial_\nu\partial^\nu\partial_\rho h^{\rho\sigma}=0
\label{26}
\end{equation}
in cartesian coordinates, or

\begin{equation}
\partial_\mu\partial_\nu (\partial^\sigma B^\nu-\partial^\nu B^\sigma)=0\,,
\label{27}
\end{equation}
where $B^\mu=\partial_\rho h^{\rho\mu}$. The field equation above is invariant
under the gauge transformation

\begin{equation}
B^\mu \rightarrow B^\mu + \partial^\mu \Lambda\,,
\label{28}
\end{equation}
where $\Lambda$ is an arbitrary space-time dependent scalar function. The
transformation for the field $h^{\mu\nu}$ that yields Eq. (28) is given
by $h^{\mu\nu} \rightarrow h^{\mu\nu} + \eta^{\mu\nu} \Lambda$. We may 
impose a  Lorentz type gauge condition,

\begin{equation}
\partial_\mu B^\mu=0\,,
\label{29}
\end{equation}
after which the field equation (27) is reduced to 

\begin{equation}
\partial_\mu (\partial_\nu \partial^\nu) B^\sigma=0\,.
\label{30}
\end{equation}

The Newtonian limit for the gravitational field is characterized by line
element 

\begin{equation}
ds^2=-\biggl( 1+{{2\phi}\over {c^2}}\biggr)dt^2+
\biggl(1-{{2\phi}\over {c^2}}\biggr)
(dx^2 + dy^2 + dz^2)\,,
\label{31}
\end{equation}
or

\begin{equation}
h_{00}=h_{11}=h_{22}=h_{33}=-{{2\phi}\over {c^2}} \,.
\label{32}
\end{equation}
In terms of $h_{\mu\nu}$, the
gauge condition (29) reads $\partial_\nu\partial_\rho h^{\rho\nu}=0$. Since
$h_{\mu\nu}$ is time-independent, the gauge condition in the Newtonian limit
reduces to

\begin{equation}
\partial_1 \partial_1 h_{11}+\partial_2 \partial_2 h_{22}+
\partial_3 \partial_3 h_{33}= 0 \,.
\label{33}
\end{equation}
Transforming now to spherical coordinates, and taking into account (32), we 
have

\begin{equation}
\nabla^2 \phi={1\over r^2}{\partial\over {\partial r}}\biggl( r^2
{{\partial \phi}\over {\partial r}}\biggr)=0\,.
\label{34}
\end{equation}
The solution of this equation is given by 

\begin{equation}
\phi(r)=a+ {b\over r}\,,
\label{35}
\end{equation}
where $a$ and $b$ are constants. In order to comply with the Newtonian limit,
we must have $a=0$ and $b=-mc^2=-MG$ (recall that $m=MG/c^2$, where $M$ is 
identified with the mass of the source and G is the gravitational constant).

The solution (35) to the gauge condition (29) is also a solution of the
field equation (30), and ultimately of the field equation (26). Note that in 
cartesian coordinates, the field equation (30) may be rewritten as 

\begin{equation}
\partial_\mu \partial_\rho (\partial_\nu \partial^\nu) h^{\rho\sigma}=0\,.
\label{36}
\end{equation}

Thus we conclude that the weak field limit of the theory determined by  eqs. 
(17), (25) and (26) allows the description of the usual Newtonian limit of 
general relativity, namely, it yields the Newtonian potential $\phi=-MG/ r$ 
provided  we identify $b=-mc^2$ in Eq. (35). Out of all possible spherically 
symmetric solutions, Eq. (35) arises from the imposition of the weak field 
approximation. However, it is likely that the time independent expression (35)
is not the only solution to eqs. (26) and (27).

\section{Final remarks}
In this article we have addressed a theory that has three distinctive 
features: (i) it is invariant under conformal transformations,  (ii) it is 
functionally similar to Maxwell's theory of electromagnetism, and (iii)
all static, spherically symmetric geometries (including the Schwarzschild 
space-time) are solutions of the vacuum field
equations. Moreover, the linearized (weak field) equations allow the 
description of the Newtonian limit of general relativity. 

The feature regarding the existence of an infinite number of solutions with
static spherical geometry is, to some extent, similar to the existence of an 
infinity of solutions to the Laplace equation in electrostatics, in the absence
of boundary conditions for the scalar potential. After imposing boundary 
conditions, one arrives at a particular solution for the physical 
configuration. In similarity to the problems in electrostatics, it may be 
possible to obtain the Schwarzschild metric by making use of the conformal 
gauge symmetry, and requiring, for instance, that the spherical surface 
determined by $r=2m$ is a null surface, as in a boundary vale problem. 
This issue will be investigated elsewhere. It is possible that the problem
of finding stationary solutions to the field equation (17) may be reduced to a 
boundary value problem. The Schwarzschild metric displays an essential 
singularity at $r=0$, and the fact that the vacuum solutions 
characterized by (20-24) are not {\it a priori} singular, is a
positive feature. We will also analyze (i) the possible emergence of potentials
that modify the standard Newtonian potential both at small and large radial 
distances, and (ii) the existence of time dependent solutions, and in
particular of plane wave solutions. 

The emergence of the solution given by eqs. (23) and (24) is not surprising.
The tensor $F_{\mu\nu}$ has a structure of a rotational, and the rotational of
a vector field endowed with spherical symmetry, vanishes. Thus, it is not
possible to verify whether the Kerr metric tensor yields a tetrad field
that is a solution of Eq. (17), by means of the procedure presented here. This 
issue will also be investigated in the future.


\begin{thebibliography}{99}

\bibitem{tHooft1}
G. 't Hooft, ``Probing the small distance structure of canonical quantum
gravity using the conformal group" [arXiv:1009.0669]

\bibitem{tHooft2}
G. 't Hooft, ``The conformal constraint in canonical quantum gravity",
[arXiv:1011.0061]; Found. Phys. {\bf 41}, 1829 (2011) [arXiv:1104.4543].

\bibitem{Mannheim1}
P. D. Mannheim, Prog. Part. Nucl. Phys. {\bf 56}, 340 (2006).

\bibitem{Mannheim2}
P. D. Mannheim, ``Making the case for conformal gravity" [arXiv:1101.2186].

\bibitem{Moon2}
T. Moon, P. Oh and J. Sohn, JCAP 1011, 005 (2010) [arXiv:1002.2549].

\bibitem{Maluf1}
J. W. Maluf and F. F. Faria, Phys. Rev. D {\bf 85}, 027502 (2012).

\bibitem{Dirac}
P. A. M. Dirac, Proc. R. Soc. London {\bf A 333}, 403 (1973).

\bibitem{Utyiama}
R. Utyiama, Prog. Theor. Phys. {\bf 53}, 565 (1975).

\bibitem{Poplawski}
N. J. Poplawski, Mod. Phys. Lett A {\bf 22}, 2701 (2007); {\bf 24}, 431
(2009).

\bibitem{Maluf2}
J. W. Maluf, Gen. Rel. Grav. {\bf 19}, 57 (1987).

\bibitem{Obukhov1}
V. N. Ponomarev and Y. N. Obukhov, Gen. Rel. Grav. {\bf 14}, 309 (1982).

\bibitem{Obukhov2}
Y. N. Obukhov, Phys. Lett A {\bf 90}, 13 (1982).

\bibitem{Nieh}
H. T. Nieh, Phys. Lett A {\bf 88}, 388 (1982).

\bibitem{Dereli}
T. Dereli and R. W. Tucker, Phys. Lett B {\bf 110}, 206 (1982).

\bibitem{Hehl}
F. W. Hehl, J. D. McCrea, E. W. Mielke and Y. Ne'eman, Phys. Rep. {\bf 258},
1 (1995).

\bibitem{Hammond1}
R. T. Hammond, Gen. Rel. Grav. {\bf 20}, 813 (1988).

\bibitem{Hammond2}
R. T. Hammond, J. Math. Phys. {\bf 31}, 2221 (1990).

\end{thebibliography}
\end{document}